# Quantum Cellular Automata from Lattice Field Theories

Michael McGuigan

Information Technology Division
Brookhaven National Laboratory
Upton, NY 11973
mcguigan@bnl.gov

Abstract

We apply the methods of lattice field theories to the quantization of cellular automata. We discuss the quantization of five main categories of cellular automata: bosonic, fermionic, supersymmetric, spin and quantum dot using path integral and operator formalisms of lattice field theories. We show that the quantization of supersymmetric cellular automata is related to recently discussed string bit models of Thorn and Bergman and represents a link of cellular automata theory to fundamental physics. We discuss spin and quantum dot cellular automata for their importance in experimental realizations and their use in quantum computation. Previous studies of quantum cellular automata utilize the wave function values as cell contents and the discretized linear Dirac equation as an update equation. We show that our approach to the quantization of fermionic cellular automata includes this utilization as a field equation, and in addition allows for nonlinearity through lattice field interactions.

## I Introduction

Cellular automata are models based on simple rules which upon deterministic time evolution exhibit surprisingly complex behavior. They have far ranging applications from traffic flow, lattice gases to classical computation [1,2,3]. Being deterministic, cellular automata are largely restricted to the description of classical systems. As most modern physical theories are based on the rules of quantum mechanics it is desirable to have rules for cellular automata that obey quantum principles. This is particularly true if one takes a fundamental viewpoint concerning the relationship of cellular automata to nature. Quantum cellular automata (QCA) are quantized versions of cellular automata theories whose cell contents have a non deterministic time evolution. They have found applications in quantum lattice gases [4,5] and universal quantum computation [6,7].

Some of the earliest discussion related to quantum cellular automata relates to the creation of physical models of quantum computation [8,9]. In [10] Grossing and Zeilinger et al discussed quantum cellular automata from the point of view of complex valued cell contents containing the probability amplitudes effectively obtaining a discrete Schrodinger equation linear in the cell values. In [11] Watrous introduced quantum cellular automata as a new quantum computational model and showed that a quantum Turing machine can be efficiently simulated on a one dimensional quantum cellular automaton. In [12] Durr and Santha give an efficient algorithm to decide if a linear quantum cellular automaton is unitary.

Like classical cellular automata, quantum cellular automata are defined with discrete space, discrete time, and cell contents taking values in a target space that can be discrete or continuous. The main difficulty in combining quantum mechanics with discrete space involves the problem species doubling of lattice fermions [13,14,15] which we will discuss in a later section along with fermionic quantum cellular automata. Combing quantum mechanics with discrete time introduces several subtle issues associated with unitarity and conserved quantities. In [16] T.D. Lee introduced discrete time in quantum mechanics through a random temporal lattice, and by treating the temporal spacing as a dynamical variable was able to derive energy conservation. In [17] Verstegen discussed the quantum harmonic oscillator with discrete time on a regular spaced temporal lattice and found that actions with the correct classical limit do not necessarily have the correct quantum behavior. In [18] Bender et al discuss the quantization of discret time lattice systems using the finite element method. In [19] Khorrami presented a discrete time formalism of quantum mechanics based on the Feynman path integral and showed that unitarity places restrictions on the form of the action. In [20] Jaroszkiewicz and Norton discuss the principles used for discrete time quantum mechanics and apply the formalism to the discrete time harmonic oscillator. Quantum mechanical models with discrete target space have found applications in statistical physics, string and particle theory. In a series of papers Kostov et al [21] describe the propagation of strings with discrete target space. In [22] and [23] the case of propagation of particles in discrete target space is also considered with application to the problem of a quantum walk. In section II we show that this case is related to the transition amplitude for the evolution of a cellular automata with a single bosonic cell.

Prior work on QCA have found difficulties in defining nonlinear versions of the automata's defining equations. For this reason very few QCA have been found [24]. This differs drastically from classical cellular automata where numerous models are known and classified based upon dimension and the range of interaction with neighboring cells. For quantum lattice field theories nonlinear field equations are naturally interpreted by taking the field as a dynamical variable and allowing self and nearest neighbor interactions. Thus the lattice field approach has formed a powerful approach to the problem of cellular automata. In [25] 't Hooft, Isler and Kalitzin cellular automata in 1+1 dimensions were related to quantum field theories containing Dirac fermions including interactions. In [26] Svozil investigated whether cellular automata could be used as tesselation of the quantum field and discussed the implication of the problem of fermion species doubling. In [27] Kostin described a scheme for cellular automata that describe the discrete evolution of the Schrodinger and Dirac equation and conserve probability. In [28] unitary cellular automata on a cubic lattice were used to model discrete evolution of wave functions for spinning particles. In [29] Russo describes a cellular string theory whose world sheet variables evolve under deterministic rules.

In this paper we restrict ourselves to reversible cellular automata as these have the clearest physics based description. We use the path integral formalism to promote the defining cellular automata equation to a quantized theory as this is the most straightforward quantization method with discrete time. We introduce linear and nonlinear interactions with neighboring cells by the methods of lattice field theory. We then analyze partition function and transition amplitudes of the resulng models.

This paper is organized as follows. In section I we give a basic introduction of QCA and lattice field theories. In section II we describe bosonic quantum cellular automata. We describe 1-cell bosonic QCA as a simple example related to the discrete time harmonic oscillator. We describe an M-cell bosonic QCA and relate it to a lattice version of a bosonic string. We also discuss the case of discrete target space for QCA and relate these to discrete statistical models.In section III we describe fermionic cellular automata. We relate fermionic QCA to linear and nonlinear fermion lattice field theories in 1+1 dimensions. In section IV we combine bosonic and fermionic QCA and describe supersymmetric QCA. We describe 1-cell and M-cell supersymmetric QCA and relate them to a lattice version of a supersymmetric string. In section V we describe spin cellular automata and quantum dot cellular automata as these are closely related to current experimental efforts, and relate these to fermionic QCA. In section VI we discuss the main conclusions of the paper.

**II Bosonic Quantum Cellular Automata**

Cellular automata consist of a row of cells whose contents can be updated to the next time step according to a definite rule associated to the contents of the cell and it's neighbors. These can be generalized to higher dimensions by placing the cell contents on a higher dimensional lattice. Reversible cellular automata [30,31,32] are a subclass of cellular automata that exhibit physical behavior such as locality and microscopic reversiblity. As quantum mechanics is a physical theory describing atoms, nuclei, particles and fields, the physical nature of reversible cellular automata make an excellent starting point for constructing quantum cellular automata. Also reversible cellular automata are generated by second order rules and these second order equations are related to discretized bosonic field equations, like the lattice Klein-Gordon equation, which are second order.

In [30] Margolus defines reversible cellular automata through the equation

$$X(I+1,J) = f(X(I,J-1), X(I,J), X(I,J+1)) - X(I-1,J)$$

(2.1)

Where $I$ and $J$ are discrete coordinates corresponding to the time and lattice position of each cell of the automata and $X$ takes its values in a target space $\mathbb{T}$ which gives the contents of the cell at time $I$ and position $J$. The target space $\mathbb{T}$ is usually taken to be discrete $\mathbb{Z}_k$ with the difference in (2.1) taken mod $k$; however, continuous values can also be chosen [3, pp 155-160]. Here we shall take $\mathbb{T}$ to be the real numbers $\mathbb{R}$ and return to the discrete target space at the end of the section. The function $f(X)$ determines the rule used to update the cell contents to the next time step once the values of the first two time steps are specified.

Equation (2.1.) is reversible in the sense that

$$X(I-1, J) = f(X(I, J-1), X(I, J), X(I, J+1)) - X(I+1, J)$$

(2.2)

This means that the same update function can be used to run the cellular automaton backward in time from the final two time steps arriving at the initial state of the automaton described by (2.1).

To see the physical nature of the automaton described by (2.1) consider the continuous time limit which can be obtained by introducing a lattice spacing $I = t/a_0$ taking the limit $a_0$ to zero so that $t$ becomes a continuous variable. For the simple case of one cell the index $J$ can be suppressed and we have the expansion:

$$X(t+a_0) + X(t-a_0) = a_0^2 \ddot{X}(t) + 2X(t) + \ldots$$

(2.3)

So that the update equation becomes:

$$\ddot{X}(t) = f(X(t)) - 2X(t)$$

(2.4)

The reversible structure of (2.1) has lead to the physical equation for the dynamical variable $X$ driven by a force term given by $F(X) = f(X) - 2X$.

To quantize the cellular automata we use the path integral formulation. Most studies of quantum cellular automata use the Schrodinger picture and discuss discrete evolution of the Schrodinger equation. We use the path integral formulation because it is widely used in quantum lattice field theories and it is the relationship between quantum cellular automata and lattice field theories that we wish to exploit. Also the path integral formalism represents the most straightforward approach to the problem of quantum mechanics with discrete time.

(i) One cell

To begin the discussion of quantization we again consider the simple example of a single cell so that the cell lattice position $J$ can be suppressed. For the choice of the update function $f(X) = 2X - W^2 X$ equation (2.1) becomes the discrete time harmonic oscillator equation [20,33]

$$X(I+1) = 2X(I) - W^2 X(I) - X(I-1)$$

(2.5)

This equation has solutions of the form $X(I) = e^{-ia_0\omega I} A + e^{ia_0\omega I} A^*$ for complex parameter $A$ with $\omega$ defined by $\frac{1}{2}W^2 = (1-\cos(a_0\omega))$ and $a_0$ used to scale the time lattice spacing..

To quantize the cellular automaton we introduce the action functional $S$ whose variation with respect to $X(I)$ yields the update equation (2.5). The action $S$ is therefore given by:

$$S = \sum_{I=0}^{N-1} (\frac{1}{2}(X(I+1) - X(I))^2 - \frac{1}{2}W^2 X^2(I))$$

(2.6)

The quantum transition function from initial cell contents $X(0)$ to final cell contents $X(N)$ at time step $N$ is given in the path integral formulation by:

$$K(X(N), N; X(0), 0) = \prod_{I=1}^{N-1} \int dX(I) e^{iS\{X\}}$$

(2.7)

Because the action functional $S$ is quadratic in $X$, the transition function can be evaluated exactly. We find the following result:

$$K(X_N, N; X_0, 0) = \sqrt{\frac{\sin(a_0\omega)}{2\pi i \sin(a_0\omega N)}} \exp\left(\frac{i\sin(a_0\omega)}{2\sin(a_0\omega N)}\left((X_N^2 + X_0^2)\cos(a_0\omega N) - 2X_0 X_N\right)\right)$$

(2.8)

In the limit $a_0 \to 0$ with infinite time steps and $a_0 N$ equal to the elapsed time, formula (2.8) reduces to the usual expression for the transition amplitude for the harmonic oscillator [34], which is a good check of our formalism.

(ii) $M$-cell

The M-cell case cellular automata is much more interesting because of the interaction with neighboring cells. For the $M$-cell case we reintroduce the $J$ coordinate where $J = 1, 2, \ldots, M$ and define the cellular automata from the update rule:

$$X(I+1, J) = f(X(I, J-1), X(I, J), X(I, J+1)) - X(I-1, J)$$

(2.9)

For a simple case we chose the function to be:

$$f(X(I,J-1), X(I,J), X(I,J+1)) = X(I,J-1) + X(I,J+1)$$
(2.10)

The update equation becomes:

$$X(I+1,J) = X(I,J-1) + X(I,J+1) - X(I-1,J)$$
(2.11)

This equation has classical solutions of the form $X(I,J) = X_L(I+J) + X_R(I-J)$. It is interesting that the discrete equation has the same form of solutions as the continuous case and this is related to the underlying conformal invariance of the equation [35]. Figure 1 shows the classical evolution of this cellular automaton.

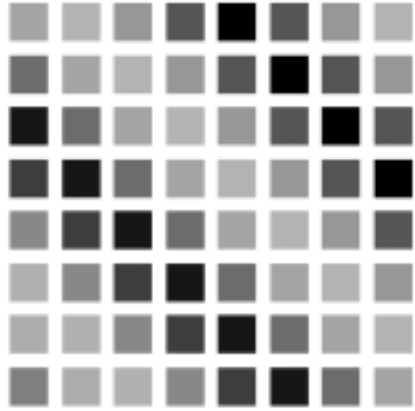

**Figure 1.** Classical evolution of the cellular automaton defined by the update rule: $X(I+1,J) = X(I,J-1) + X(I,J+1) - X(I-1,J)$. Discrete time *I* increases down and discrete space *J* across. The values of the cell contents are light gray for small values and black for large values.

To quantize the model we introduce the action $S$ whose variation with respect to $X(I,J)$ yields the update equation. Such an action $S$ is given by:

$$S = \sum_{I,J=0}^{N,M} \frac{1}{2}((X(I+1,J) - X(I,J))^2 - (X(I,J+1) - X(I,J))^2)$$
(2.12)

The transition function for the amplitude to go from cell contents $X(0,\cdot)$ to cell contents $X(N,\cdot)$ ( the $(\cdot)$ notation indicates the many *J* values) is given in the path integral formulation as:

$$K(X(N,\cdot),N;X(0,\cdot),0) = \prod_{I,J=1,0}^{N-1,M} \int dX(I,J) e^{iS\{X\}}$$

(2.13)

This model has been solved in the context of discrete string theory by Giles and Thorn [36] by Fourier transforming with respect to the $J$ coordinate through:

$$X(I,J) = \frac{1}{\sqrt{M}} \sum_{n=0}^{M-1} X_n(I) e^{-2\pi i n J/M}$$

(2.14)

The update equation then becomes:

$$X_n(I+1) = 2X_n(I) - W_n^2 X_n(I) - X_n(I-1)$$

(2.15)

with $W_n = 2\sin(\pi n/M) = 2\sin(a_0 \omega_n/2)$. The transition amplitude can then be constructed as a product of transition amplitudes of the discrete Harmonic oscillator cellular automaton obtained from (2.8).

(iii) $M$-cell Discrete Target Space

Traditionally cellular automata are defined with discrete target space $\mathbb{T}$ as well as discrete space cell location and discrete time. One class of discrete target space is cyclic group $\mathbb{Z}_k$ with $k$ elements $0,1,\ldots,k-1$ with arithmetic defined modulo $k$. Transition amplitudes defined on such a space should be periodic modulo $k$ that is $X \in \mathbb{Z}_k$, $X \approx X + k$ with:

$$K(X(N,\cdot)+k,N;X(0,\cdot),0) = K(X(N,\cdot),N;X(0,\cdot)+k,0) = K(X(N,\cdot),N;X(0,\cdot),0)$$

(2.16)

One way to enforce this condition is to define the model on an infinite target space $\mathbb{Z}$ and then introduce the infinite sum

$$K(X(N,\cdot),N;X(0,\cdot),0) = \sum_{m=-\infty}^{\infty} \sum_{X(I,J)=0}^{k-1} e^{iS\{X+mk\}}$$

(2.17)

For example in the transition amplitude in continuous time for a quantum walk on $\mathbb{Z}_k$ [23] was obtained from the quantum walk on $\mathbb{Z}$ through:

$$K_{\mathbb{Z}_k}(X_f,t_f;X_i,t_i) = \sum_{m=-\infty}^{\infty} K_{\mathbb{Z}}(X_f,t_f;X_i+mk,t_i) = \sum_{m=-\infty}^{\infty} \frac{1}{2\pi} \int_{-\pi}^{\pi} dp\, e^{ip(X_i+mk-X_f)} e^{-2i(t_f-t_i)(\cos(p)-1)}$$

(2.18)

Another method to impose a discrete target space is to introduce target space fields $Y(I,J), Z(I,J)$ and impose the constraint $Y^2(I,J)+Z^2(I,J)=1$ through the parameterization:

$$(Y(I,J),Z(I,J)) = \frac{k}{2\pi}(\cos(2\pi X(I,J)/k),\sin(2\pi(X(I,J)/k)))$$

(2.19)

The periodic relation will then be imposed directly by the trigonometric functions. For example for the update functions
$f(Y(I,J-1),Y(I,J),Y(I,J+1)) = Y(I,J-1)+Y(I,J+1)$ and
$f(Z(I,J-1),Z(I,J),Z(I,J+1)) = Z(I,J-1)+Z(I,J+1)$ yield rules which follow from the action:

$$S = \sum_{I,J=0}^{N-1,M} \frac{1}{2}((Y(I+1,J)-Y(I,J))^2 - (Y(I,J+1)-Y(I,J))^2 + (Z(I+1,J)-Z(I,J))^2 - (Z(I,J+1)-Z(I,J))^2)$$

(2.20)

In terms of the parameterization (2.19) the action becomes

$$S = \left(\frac{k}{2\pi}\right)^2 \sum_{I,J=0}^{N-1,M} \cos(2\pi(X(I,J+1)-X(I,J))/k) - \cos(2\pi(X(I+1,J)-X(I,J))/k)$$

(2.21)

with transition amplitude

$$K_{\mathbb{Z}_k}(X(N,\cdot),N;X(0,\cdot)) = \prod_{I,J=1,0}^{N-1,M} \sum_{X(I,J)=0}^{k-1} e^{iS\{X\}}$$

(2.22)

Note that for the discrete target space the integral over $X$ reduces to a finite sum and the form of the action (2.21) becomes a discrete statistical mechanical model upon rotation to Euclidean space.

One further special case of interest is the $\mathbb{Z}_2$ target space. This is the most popular target space in classical cellular automata theory where each cell is dead or alive taking the value $0$ or $1$. The action in this case becomes:

$$S = \left(\frac{1}{\pi}\right)^2 \sum_{I,J=0}^{N-1,M} \cos(\pi X(I,J+1))\cos(\pi X(I,J)) - \cos(\pi X(I+1,J))\cos(\pi X(I,J))$$
(2.23)

Defining the spin variables $S_z(I,J) = \cos(\pi X(I,J))$ the action becomes

$$A = \left(\frac{1}{\pi}\right)^2 \sum_{I,J=0}^{N-1,M} S_z(I,J+1)S_z(I,J) - S_z(I+1,J)S_z(I,J)$$
(2.24)

where we switch the notation for the action to $A$ so as to avoid confusion with the spin variables. The transition amplitude is then

$$K(S_z(N,\cdot), N; S_z(0,\cdot)) = \prod_{I,J=0}^{N-1,M} \sum_{S_z(I,J)=\pm 1} e^{iA\{S_z\}}$$
(2.25)

and is related to the statistical mechanical Ising model upon rotation to Euclidean space. More relations between Euclidean path integrals and statistical mechanics can be found in the work of Creutz and Freedman [33].

**III Fermionic Quantum Cellular Automata**

A natural generalization from the bosonic cellular automata discussed in section II is fermionic cellular automata. This is important particularly for cellular models of condensed matter or fundamental models like string theory or quark lattice gauge theory where basic degrees of freedom are fermionic. Again we shall work with reversible cellular automata with the main variation that the updated cell contents are anticommuting. $\theta(I,J)$. The defining update equation is given by:

$$\theta(I+1,J) = f(\theta(I,J-1), \theta(I,J), \theta(I,J+1)) + \theta(I-1,J)$$
(3.1)

The plus sign on the right hand side is important to obtain a equation linear in time derivatives when the timelike lattice spacing is take to zero; however, the equation still remains reversible.

(i)    One cell case.

To begin consider the one cell case. In the one cell case we suppress the variable J that denotes cell location and the update equation becomes:

$$\theta(I+1) = f(\theta(I)) + \theta(I-1)$$

(3.2)

For the update function

$$f(\theta(I)) = -i2W\theta(I)$$

(3.3)

The cellular automaton represents a discrete time version of the fermionic simple harmonic oscillator.

$$\theta(I+1) = -i2W\theta(I) + \theta(I-1)$$

(3.4)

The update equation has the classical solution $\theta(I) = be^{-ia_0\omega I}$ if we parametrize $W = \sin(a_0 w)$. To quantize the cellular automaton we seek a action which yields the update equation upon variation with respect to $\theta(I)$. A suitable action is the following:

$$S = \sum_{I=0}^{N-1} i\theta^*(I)(\frac{1}{2}(\theta(I+1) - \theta(I-1))) - W\theta^*(I)\theta(I)$$

(3.5)

The action in the present form however suffers from the well known problem of fermion doubling [13] where the inverse propagator possesses extra zeros which are not present in a continuum treatment. To see this Fourier transform the $\theta(I)$ to obtain the inverse propagator $D_F(p_0) = \sin(p_0)$ with zeros at $0$ and $\pi$. One solution to the fermion doubling problem is to introduce a Wilson term into the action with parameter $r$ [37]. The action then becomes:

$$S_r = \sum_{I=0}^{N-1} i\theta^*(I)(\frac{1}{2}(\theta(I+1) - \theta(I-1)) - \frac{r}{2}(\theta(I+1) + \theta(I-1) - 2\theta(I))) - W\theta^*(I)\theta(I)$$

(3.6)

The inverse propagator after Fourier transform now becomes $D_F(p_0) = \sin(p_0) + r(1 - \cos(p_0)) = \sin(p_0) + 2r\sin^2(p_0/2)$ a form with a single zero at $p_0 = 0$.

To quantize the model we again go to the path integral formulation with discrete time. The path integral for $N$ time steps is given by integral over Grassmann variables $\theta(I)$ in the following form:

$$K(\theta(N), N; \theta(0), 0) = \prod_{I=1}^{N-1} d\theta(I) d\theta^*(I) e^{iS_r\{\theta\}}$$

(3.7)

(ii)     Fermionic $M$-cell case

The $M$-cell case can be studied similarly. The simplest choice for the function update function is $f(\theta(I,J-1), \theta(I,J), \theta(I,J+1)) = \theta(I,J-1) - \theta(I,J+1)$ so that update equation becomes

$$\theta(I+1, J) = \theta(I, J-1) - \theta(I, J+1) + \theta(I-1, J)$$

(3.8)

The classical solution to the update equation is of the form $\theta(I,J) = \theta_R(I-J)$ and is the same form as the continuum solution a result indicative of an underlying conformal invariance [35]. Introducing a second fermionic field $\tilde{\theta}(I,J)$ with update function $\tilde{f}(\tilde{\theta}(I,J-1), \tilde{\theta}(I,J), \tilde{\theta}(I,J+1)) = -\tilde{\theta}(I,J-1) + \tilde{\theta}(I,J+1)$ and update equation

$$\tilde{\theta}(I+1, J) = -\tilde{\theta}(I, J-1) + \tilde{\theta}(I, J+1) + \tilde{\theta}(I-1, J)$$

(3.9)

This equation has classical solutions of the form $\tilde{\theta}(I,J) = \tilde{\theta}_L(I+J)$.

To quantize the fermionic cellular automata we form a doublet $\Theta_\alpha = \begin{pmatrix} \theta \\ \tilde{\theta} \end{pmatrix}$ with two dimensional representation of the Dirac matrices $\rho^0 = \begin{pmatrix} 0 & -i \\ i & 0 \end{pmatrix}$ and $\rho^1 = \begin{pmatrix} 0 & i \\ i & 0 \end{pmatrix}$. The Dirac action $S_D = \int d^2\sigma (-i\overline{\Theta}_a (\rho^\mu \partial_\mu)^{ab} \Theta_b)$ in discretized form writing out the components we have:

$$S = \sum_{I,J=0}^{N-1, M-1} -i\frac{1}{2}\theta(I,J)(\theta(I+1,J) - \theta(I-1,J)) - i\frac{1}{2}\theta(I,J)(\theta(I,J+1) - \theta(I,J-1))$$
$$-i\frac{1}{2}\tilde{\theta}(I,J)(\tilde{\theta}(I+1,J) - \tilde{\theta}(I-1,J)) + i\frac{1}{2}\tilde{\theta}(I,J)(\tilde{\theta}(I,J+1) - \tilde{\theta}(I,J-1))$$

(3.10)

Variation of this action clearly leads to the update equation (3.8) and (3.9).

The action again suffers from the fermion doubling problem as can be see by Fourier transformation where the inverse propagator is given by $D_F(p_0, p_1) = \sin(p_0) + \sin(p_1)$ with zeros located at $(0,0), (0,\pi), (\pi,0), (\pi,\pi)$.

The introduction of a Wilson term with parameter $r$ again removes the extra doublers however in this case the action is not chiral invariant with solutions that are not purely right moving. The action with the Wilson term is given by:

$$S_r = \sum_{I,J=0}^{N-1,M-1} -i\frac{1}{2}(\theta(I,J)(\theta(I+1,J) - \theta(I-1,J) - r\tilde{\theta}(I+1,J) - r\tilde{\theta}(I-1,J) + 2r\tilde{\theta}(I,J))$$
$$-i\frac{1}{2}(\theta(I,J)(\theta(I,J+1) - \theta(I,J-1) - r\tilde{\theta}(I,J+1) - r\tilde{\theta}(I,J-1) + 2r\tilde{\theta}(I,J))$$
$$-i\frac{1}{2}(\tilde{\theta}(I,J)(\tilde{\theta}(I+1,J) - \tilde{\theta}(I-1,J) + r\theta(I+1,J) + r\theta(I-1,J) - 2r\theta(I,J))$$
$$+i\frac{1}{2}(\tilde{\theta}(I,J)(\tilde{\theta}(I,J+1) - \tilde{\theta}(I,J-1) - r\theta(I,J+1) - r\theta(I,J-1) + 2r\theta(I,J))$$

(3.11)

In this form the inverse propagator has a single zero at $(p_0, p_1) = (0,0)$.

Finally to quantize this system we again use the Path integral formulation to compute the transition amplitude given by:

$$K(\theta(N,\cdot), N; \theta(0,\cdot)) = \prod_{I,J=1,0}^{N-1,M} \int d\theta(I,J) d\tilde{\theta}(I,J) e^{iS_r\{\theta\}}$$

(3.12)

(iii)    Nonlinear case

Choosing a nonlinear update function will lead to a nonlinear cellular automata. For example a choice of the form $f(\theta) = g\theta\rho^\alpha \bar{\theta}\rho_\alpha \theta$ will lead to a cellular automata associated to the discretization of the Thirring model [38-42] with four-fermion interaction with $\rho^\alpha$ the two dimensional dirac matrices. Another choice for nonlinear update function is:

$$f(\theta(I,J-1), \theta(I,J), \theta(I,J+1)) = \theta(I,J+1) - \theta(I,J-1)$$
$$+ g\theta(I,J)\theta^*(I,J+1)\theta(I,J+1) + g\theta(I,J-1)\theta^*(I,J)\theta(I,J-1)$$

(3.13)

This form of the update function yields a four-fermion action bosonizes into the anisotropic Heisenberg Spin model and was studied by Creutz [43] as a non-trivial example of the evaluation of Grassmann integrals by machine. We will encounter further nonlinear fermionic cellular automata in later sections in relation to Spin and Quantum-Dot cellular automata as well as universal control Hamiltonians for quantum computing.

**IV Supersymmetric Quantum Cellular Automata**

Supersymmetric cellular automata results from combing the above results on bosonic and fermionic cellular automata with the introduction of a discrete supersymmetry (SUSY) relating the bosonic and fermionic cell contents. These models can be used for studies of supersymmetric lattice field theories [44-48], supersymmetric condensed matter theories such as the supersymmetric Hubbard or TJ model [49,50], and as a discrete formulation of superstring theory [36,51-57]. The introduction of bosonic and fermionic update equations with two update functions $f_1$ and $f_2$

$$X(I+1,J) = f_1(X(I,J-1), \theta(I,J-1), X(I,J), \theta(I,J), X(I,J+1), \theta(I,J+1)) - X(I,J-1)$$
$$\theta(I+1,J) = f_2(X(I,J-1), \theta(I,J-1), X(I,J), \theta(I,J), X(I,J+1), \theta(I,J+1)) + \theta(I,J-1)$$
(4.1)

However the minus on the right for the update equation for $X$ means that there will be no doubling for $X$. While the plus on the right for the update equation for $\theta$ will lead to doubling. This will present a barrier to supersymmetry on discrete space and time. One way to address this is to introduce a first order formalism for the bosons so that both bosons and fermions have doublers:

$$P(I+1,J) = f_1(X(I,J-1), \theta(I,J-1), X(I,J), \theta(I,J), X(I,J+1), \theta(I,J+1)) + P(I,J-1)$$
$$X(I+1,J) = 2P(I,J) + X(I-1,J)$$
$$\theta(I+1,J) = f_2(X(I,J-1), \theta(I,J-1), X(I,J), \theta(I,J), X(I,J+1), \theta(I,J+1)) + \theta(I,J-1)$$
(4.2)

and then add Wilson terms to remove the doublers in a manner that preserves supersymmetry.

We consider examples of this method in the 1-cell and M-cell supersymmetric automata.

(i)   One cell case

For the simplest case of one cell we can suppress the J index. The easiest case to consider is the discrete supersymmetric simple harmonic oscillator [45] with update equations:

$$P(I+1) = -2W^2 X(I) + P(I-1)$$
$$X(I+1) = 2P(I) + X(I-1) \qquad (4.3)$$
$$\theta(I+1) = -2iW\theta(I) + \theta(I-1)$$

The action which generates these update equations is given by:

$$S = \sum_{I=1}^{N-1} i\bar{\theta}(\frac{1}{2}(\theta(I+1)-\theta(I-1))) - W\bar{\theta}(I)\theta(I) + P(I)(\frac{1}{2}(X(I+1)-X(I-1))) - \frac{1}{2}P^2(I) - \frac{W^2}{2}X^2(I)$$
(4.4)

Then the supersymmetric transformations that leave the action invariant are

$$\delta P = \bar{\xi}\frac{1}{2}(\theta(I+1)-\theta(I-1))$$
$$\delta X = \bar{\xi}\theta(I)$$
$$\delta\bar{\theta} = \bar{\xi}i\frac{1}{2}(X(I+1)-X(I-1)) - W\bar{\xi}X(I)$$
$$\delta\theta = 0$$
(4.5)

As we mentioned before the present form of the action and SUSY transformations have doublers for bosonic and fermionic fields. Adding Wilson terms for the bosons and fermions amounts to the substitution:

$$X(I+1) - X(I-1) \Rightarrow X(I+1) - X(I-1,J) - r(X(I+1) + X(I-1) - 2X(I))$$
$$\theta(I+1) - \theta(I-1) \Rightarrow \theta(I+1) - \theta(I-1) - r(\theta(I+1) + \theta(I-1) - 2\theta(I))$$
(4.6)

The action then becomes

$$S_r = \sum_{I=1}^{N-1} i\bar{\theta}(\frac{1}{2}(\theta(I+1)-\theta(I-1)) - \frac{r}{2}(\theta(I+1)+\theta(I-1)-2\theta(I))) - W\bar{\theta}(I)\theta(I) +$$
$$\frac{1}{2}P(I)(X(I+1)-X(I-1) - \frac{r}{2}(X(I+1)+X(I-1)-2X(I))) - \frac{1}{2}P^2(I) - \frac{W^2}{2}X^2(I)$$
(4.7)

and supersymmetry transformations with Grassmann parameter $\bar{\xi}$ are

$$\delta P = \bar{\xi}\frac{1}{2}(\theta(I+1)-\theta(I-1) - r(\theta(I+1)+\theta(I-1)-2\theta(I)))$$
$$\delta X = \bar{\xi}\theta(I)$$
$$\delta\bar{\theta} = \bar{\xi}i\frac{1}{2}(X(I+1)-X(I-1) - r(X(I+1)+X(I-1)-2X(I))) - W\bar{\xi}X(I)$$
$$\delta\theta = 0$$
(4.8)

Finally to quantize we can again use the discrete time path integral approach defined by

$$K(X(N),\theta(N),\bar{\theta}(N)N;X(0),\theta(0),\bar{\theta}(0)) = \prod_{I=1}^{N-1}\int dX(I)d\theta(I)d\bar{\theta}(I)e^{iS_r\{X,\theta,\bar{\theta}\}}$$

(4.9)

(i)     M-cell case

The simplest supersymmetric M-cell case is obtained by combing (2.11) and (3.8). In the first order formalism the action is given by:

$$S = \sum_{I,J=0}^{N-1,M-1} \frac{1}{2}(-P(I,J)(X(I+1,J)-X(I-1,J))+L(I,J)(X(I,J+1)-X(I,J-1))+P(I,J)^2-L(I,J)^2)$$

$$+i\frac{1}{2}\theta(I,J)(\theta(I+1,J)-\theta(I-1,J))+i\frac{1}{2}\theta(I,J)(\theta(I,J+1)-\theta(I,J-1))$$

$$+i\frac{1}{2}\tilde{\theta}(I,J)(\tilde{\theta}(I+1,J)-\tilde{\theta}(I-1,J))-i\frac{1}{2}\tilde{\theta}(I,J)(\tilde{\theta}(I,J+1)-\tilde{\theta}(I,J-1))$$

(4.10)

and the supersymmetric transformations for Grassmann parameter $\xi$ are :

$$\delta P = i\frac{\tilde{\xi}}{2}(\theta(I+1,J)-\theta(I-1,J))-i\frac{\xi}{2}(\tilde{\theta}(I+1,J)-\tilde{\theta}(I-1,J))$$

$$\delta L = i\frac{\tilde{\xi}}{2}(\theta(I,J+1)-\theta(I,J-1))-i\frac{\xi}{2}(\tilde{\theta}(I,J+1)-\tilde{\theta}(I,J-1))$$

$$\delta X = i(\tilde{\xi}\theta(I,J)-\xi\tilde{\theta}(I,J))$$

$$\delta\theta = -\frac{\tilde{\xi}}{2}(X(I+1,J)-X(I-1,J)-X(I,J+1)+X(I,J-1))$$

$$\delta\tilde{\theta} = \frac{\xi}{2}(X(I+1,J)-X(I-1,J)+X(I,J+1)-X(I,J-1))$$

(4.11)

The action and SUSY transformations suffer from boson and fermion doubling in such a way that supersymmetry is preserved. To remove the doublers we introduce Wilson terms for both the bosons and fermions as in [45]. This amounts to the following substitution in the action $S_r$ and SUSY transformations.

$$X(I+1,J)-X(I-1,J) \Rightarrow X(I+1,J)-X(I-1,J)-r(X(I+1,J)+X(I-1,J)-2X(I,J))$$
$$X(I,J+1)-X(I,J-1) \Rightarrow X(I,J+1)-X(I,J-1)-r(X(I,J+1)+X(I,J-1)-2X(I,J))$$
$$\theta(I+1,J)-\theta(I-1,J) \Rightarrow \theta(I+1,J)-\theta(I-1,J)-r(\tilde{\theta}(I+1,J)+\tilde{\theta}(I-1,J)-2\tilde{\theta}(I,J))$$
$$\theta(I,J+1)-\theta(I,J-1) \Rightarrow \theta(I,J+1)-\theta(I,J-1)-r(\tilde{\theta}(I,J+1)+\tilde{\theta}(I,J-1)-2\tilde{\theta}(I,J))$$
$$\tilde{\theta}(I+1,J)-\tilde{\theta}(I-1,J) \Rightarrow \tilde{\theta}(I+1,J)-\tilde{\theta}(I-1,J)+r(\theta(I+1,J)+\theta(I-1,J)-2\theta(I,J))$$
$$\tilde{\theta}(I,J+1)-\tilde{\theta}(I,J-1) \Rightarrow \tilde{\theta}(I,J+1)-\tilde{\theta}(I,J-1)-r(\theta(I,J+1)+\theta(I,J-1)-2\theta(I,J))$$

(4.12)

Substitution into the action yields:

$$S_r = \sum_{I,J=0}^{N-1,M-1} \frac{1}{2}(-P(I,J)(X(I+1,J)-X(I-1,J)-r(X(I+1,J)+X(I-1,J)-2X(I,J)))$$
$$+L(I,J)(X(I,J+1)-X(I,J-1)-r(X(I,J+1)+X(I,J-1)-2X(I,J)))+P(I,J)^2-L(I,J)^2)$$
$$+i\frac{1}{2}\theta(I,J)(\theta(I+1,J)-\theta(I-1,J)-r(\tilde{\theta}(I+1,J)+\tilde{\theta}(I-1,J)-2\tilde{\theta}(I,J)))$$
$$+i\frac{1}{2}\theta(I,J)(\theta(I,J+1)-\theta(I,J-1)-r(\tilde{\theta}(I,J+1)+\tilde{\theta}(I,J-1)-2\tilde{\theta}(I,J)))$$
$$+i\frac{1}{2}\tilde{\theta}(I,J)(\tilde{\theta}(I+1,J)-\tilde{\theta}(I-1,J)+r(\theta(I+1,J)+\theta(I-1,J)-2\theta(I,J)))$$
$$-i\frac{1}{2}\tilde{\theta}(I,J)(\tilde{\theta}(I,J+1)-\tilde{\theta}(I,J-1)-r(\theta(I,J+1)+\theta(I,J-1)-2\theta(I,J)))$$

(4.13)

While substitution into the SUSY transformations gives:

$$\delta P = i\frac{\xi}{2}(\theta(I+1,J)-\theta(I-1,J)-r(\tilde{\theta}(I+1,J)+\tilde{\theta}(I-1,J)-2\tilde{\theta}(I,J)))-i\frac{\xi}{2}(\tilde{\theta}(I+1,J)-\tilde{\theta}(I-1,J)+r(\theta(I+1,J)+\theta(I-1,J)-2\theta(I,J)))$$
$$\delta L = i\frac{\xi}{2}(\theta(I,J+1)-\theta(I,J-1)-r(\tilde{\theta}(I,J+1)+\tilde{\theta}(I,J-1)-2\tilde{\theta}(I,J)))-i\frac{\xi}{2}(\tilde{\theta}(I,J+1)-\tilde{\theta}(I,J-1)-r(\theta(I,J+1)+\theta(I,J-1)-2\theta(I,J)))$$
$$\delta X = i(\xi\theta(I,J)-\xi\tilde{\theta}(I,J))$$
$$\delta\theta = -\frac{\xi}{2}(X(I+1,J)-X(I-1,J)-r(X(I+1,J)+X(I-1,J)-2X(I,J))-X(I,J+1)+X(I,J-1)+r(X(I,J+1)+X(I,J-1)-2X(I,J)))$$
$$\delta\tilde{\theta} = \frac{\xi}{2}(X(I+1,J)-X(I-1,J)-r(X(I+1,J)+X(I-1,J)-2X(I,J))+X(I,J+1)-X(I,J-1)-r(X(I,J+1)+X(I,J-1)-2X(I,J)))$$

(4.14)

The transition amplitude for M-cell supersymmetric quantum cellular automaton is then given by the path integral:

$$K(X(N,\cdot),\theta(N,\cdot),\tilde{\theta}(N,\cdot)N;X(0,\cdot),\theta(0,\cdot),\tilde{\theta}(0,\cdot)) =$$
$$\prod_{I,J=1,0}^{N-1,M}\int dX(I,J)d\theta(I,J)d\tilde{\theta}(I,J)e^{iS_r\{X,\theta,\tilde{\theta}\}}$$

(4.15)

This form of the action is directly related to string bit models of Thorn and Bergman [52-55]. String bit models are formulations of string theory in $D-1$ space dimensions as lattice field theories of point like string bits moving in $d = D-2$ space dimensions. Here the string is composed of bits forming a polymer like chain that reproduce string theory in the continuum limit. In particular the continuum limit of the action (4.13) is the Type IIB superstring with 2+1 dimensional Target space whose Lagrangian the light cone gauge is given by:

$$S = \int d^2\sigma (\partial_\alpha X \partial^\alpha X - i\bar{\Theta}\rho^\alpha \partial_\alpha \Theta)$$

(4.16)

Where the two dimensional Dirac matrices are given by:

$$\rho^0 = \begin{pmatrix} 0 & -i \\ i & 0 \end{pmatrix} \quad ; \quad \rho^1 = \begin{pmatrix} 0 & i \\ i & 0 \end{pmatrix} \quad \text{with the doublet } \Theta = \begin{pmatrix} \theta \\ \tilde{\theta} \end{pmatrix}.$$

(4.17)

For more realistic heterotic superstring models in $D=10$ spacetime dimensions one has two dimensional chiral fermions and the $\tilde{\theta}$ field is not present in the continuum limit. This is problematic in forming as superstring bit model or supersymmetric quantum cellular automata as the $\tilde{\theta}$ field is used to remove the doublers from the $\theta$ field and visa versa through the Wilson terms. In [52] Thorn and Bergman discuss the obstacles present in the chiral case. They point out that Kogut-Susskind resolution of the doubling problem which utilizes the doubler modes as part of the observable spectrum would destroy the heterotic nature of the model as the string would have left and right moving fermionic modes. They use the Wilson term but with an asymmetry in the action between left and right moving fermions proportional to a parameter $\eta$. As $\eta \to \infty$ the left moving modes receive infinite energy and decouple from the theory.

The domain wall fermion approach [58-61] to the lattice chiral fermion problem can also be applied to the heterotic string model. Again the Wilson term is employed but in one higher dimension, in this case three. The left moving modes are confined to one domain wall and the right moving modes to the other. In the continuum limit one has effectively a two dimensional theory of chiral fermions with the left moving modes confined to one wall and decoupled from the theory. One interesting aspect of this domain wall approach is that the discretization of the three dimensional theory is used to relate the M-atrix theory of Banks, Fischler, Shenker and Susskind [62] to a three dimensional membrane action through the correspondence:

$$X(I,J,K) = \sum_{n,m} X_{nm}(a_0 I) e^{2\pi i(nJ+mK)/M} \to X(t) = \sum_{n,m} X_{nm}(t) U^n V^m$$

(4.18)

Where the matrices $U$ and $V$ satisfy $UV = e^{\frac{2\pi i}{M}} VU$ so that $X(t) \in U(M)$. In this context the M-atrix theory supersymmetric quantum mechanics discussed in [62-69] together with it's discrete time generalization [70] become an example of a supersymmetric quantum cellular automata in two space and one time discrete dimensions.

## V Spin and Quantum Dot Cellular Automata

Spin systems and quantum dot cellular automata are especially interesting because of their experimental realization and application to quantum scale devices and quantum computing.

(i) Spin cellular automata

Spin cellular automata are defined with three cell contents $S_a(I,J)$ $a = x, y, z$ and defined obey the update equation

$$S_a(I+1,J) = f_a(S_b(I,J+1), S_b(I,J), S_b(I,J+1)) + S_a(I-1,J)$$
(5.1)

For the case when the update function is of the form:

$$f_a(S_b(I,J+1), S_b(I,J), S_b(I,J-1)) = 2\varepsilon_{abc} S_b(I,J) B_c + 2\varepsilon_{abc} S_b(I,J) S_c(I,J-1) + 2\varepsilon_{abc} S_b(t,J) S_c(t,J+1)$$
(5.2)

the update equation becomes:

$$S_a(I+1,J) = 2\varepsilon_{abc} S_b(I,J) B_c + 2\varepsilon_{abc} S_b(I,J) S_c(I,J-1) + 2\varepsilon_{abc} S_b(t,J) S_c(t,J+1) + S_a(I-1,J)$$
(5.3)

In the continuum time limit $t = a_0 I$, $a_0 \to 0$ we have the Bloch equation with magnetic field $B_a$

$$\partial_t S_a(t,J) = \varepsilon_{abc} S_b(t,J) B_c + \varepsilon_{abc} S_b(t,J) S_c(t,J-1) + \varepsilon_{abc} S_b(t,J) S_c(t,J+1)$$
(5.4)

For spin and quantum-dot systems it is more convenient to take the continuum time limit and quantize using a Hamiltonian. If we promote the classical variable $S_a(t,J)$ to a quantum operator, use the spin algebra $[S_a, S_b] = i\varepsilon_{abc} S_c$ and quantize in the Heisenberg picture, the equation is written

$$\partial_t S_a(t,J) = -i[S_a(t,J), H]$$
(5.5)

with

$$H = \sum_{J=1}^{M} S_a(t,J) S_a(t,J+1) + S_a(t,J) B_a$$
(5.6)

This is the Hamiltonian for the Heisenberg Spin model with nearest neighbor interactions in one dimension. and the transition amplitude is given by:

$$K(S_z(T,\cdot); S_z(0,\cdot)) = <S_z(T,\cdot)|e^{-iTH}|S_z(0,\cdot)>.$$

(5.7)

This model with $B_a = (0,0,1)$ can be mapped on to a fermion model in one spatial dimension through the Jordan Wigner transformation [71]:

$$\theta(J) = e^{i\pi \sum_{K=1}^{J-1} S_+(K)S_-(K)} S_-(J) \quad ; \quad \theta^\dagger(J) = e^{-i\pi \sum_{K=1}^{J-1} S_+(K)S_-(K)} S_+(J)$$

(5.8)

Where $S_\pm = S_x \pm iS_y$. Then the Hamiltonian can be written [39]:

$$H = \sum_{J=1}^{M} \frac{1}{2}(\theta(J)^\dagger \theta(J+1) + \theta^\dagger(J+1)\theta(J)) + \theta^\dagger(J)\theta(J)\theta^\dagger(J+1)\theta(J+1)$$

(5.9)

In this form we see the nonlinear interaction for the fermionic quantum cellular automata from section III. An alternate approach is to evaluation is to apply the Bethe ansatz [72,73] form of the wave functions for (5.6) and use these to form the transition amplitude.

(ii)     Quantum dot cellular automata

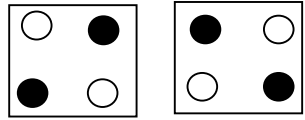

**Figure 2.** Four-site two electron quantum dot system yielding a two state system given by the left and right squares with electron locations given by dark dots.

A quantum dot cellular automata is a quantum device consisting of an array of cells with each cell content determined by an artificial crystal of several dots containing multiple electrons [74-80]. Figure 2. shows a two unit cells of a quantum dot cellular automata consisting of a four dot two electron system. This is effectively a two state system with Hamiltonian (Adachi and Isawa (1998) [78], Cole and Lusth (2001) [80]):

$$H_{eff} = \sum_{K=1}^{M} B_z S_z(K) + \sum_{K=1}^{M} \gamma S_x(K) S_x(K+1)$$

(5.10)

The model as written is equivalent to the Ising model with transverse magnetic field and can be described by a spin cellular automaton. This model is mapped through the Jordan-Wigner transformation (5.8) to a system of spinless fermions with Hamiltonian [81,82]:

$$H_{eff} = -B_z \frac{M}{2} + B_z \sum_{K=1}^{M} \theta^\dagger(K)\theta(K) + \frac{\gamma}{4} \sum_{K=1}^{M} (\theta^\dagger(K) - \theta(K))(\theta^\dagger(K+1) - \theta(K+1))$$
$$- \frac{\gamma}{4}(\theta^\dagger(M) - \theta(M))(\theta^\dagger(1) - \theta(1))(e^{\sum_{J=1}^{M} \theta^\dagger(J)\theta(J)} + 1)$$

(5.11)

For large M the last term can be neglected and one has a Hamiltonian quadratic in the fermion fields that can be solved exactly. In either form (5.10) or (5.11) the quantum dot cellular automaton is a subclass of spin and fermionic cellular automata.

(iii)   Quantum cellular automata and quantum computing

Quantum cellular automata were among the first quantum computing architectures proposed [83]. They have advantages because they require only homogeneous magnetic fields and quantum error correction at only a small number of frequencies. A scheme has been proposed for quantum cellular automata computing using global addressing techniques applied to endohedral fullerene molecules [84] as well as the quantum dot arrays discussed above. In addition to these specific proposals for quantum computation the standard models for quantum computation using control Hamiltonians can be included within the context of quantum cellular automata as we will now describe.
.
The control Hamiltonian for the standard spin model of quantum computation is given by:

$$H_{qc} = \sum_{K=1}^{M} (\alpha_x^K(t) S_x(K) + \beta_y^K(t) S_y(K)) + \sum_{K,L=1}^{M} \gamma^{KL}(t) S_z(K) S_z(L)$$

(5.12)

In the standard spin model of quantum computation the spin control Hamiltonian for one and two qubit operations is sufficient to build up any unitary operation (Barenco et al (1995) [85], Di Vincenzo (1995) [86]). In this model quantum algorithms are described by the specification of control functions $\alpha_x^K(t)$, $\beta_y^K(t)$, $\gamma^{KL}(t)$. The Jordan-Wigner transformation is used to map the spin control Hamiltonian to the the control Hamiltonian for the Standard Fermionic model of quantum computation. (Ortiz, Gubernatis, Knill, Laflamme (2000) [87] which is given by:

$$H_{qc} = \sum_{K=1}^{M} (a_K(t)\theta(K) + b_K(t)\theta^\dagger(K)) + \sum_{K,L=1}^{M} w_{KL}(t)(\theta^\dagger(K)\theta(L) + \theta^\dagger(L)\theta(K)) + g_{KL}(t)\theta^\dagger(K)\theta(K)\theta^\dagger(L)\theta(L)$$

(5.13)

Again in either form the spin and fermionic standard model control Hamiltonian for quantum computation are a subclass of spin and fermionic quantum cellular automata.

One of the most interesting features of classical cellular automata is that complex patterns of cells can emerge from very simple update rules. One way to see that classical cellular automata are capable of complex behavior is that classical cellular automata include computers, and computers are capable of simulating complex systems. As quantum cellular automata include quantum computers they are also capable simulating complex systems. However in this case these systems are of nondeterministic or probablistic character from underlying Lagrangians with simple structure and nearest neighbor interactions.

Another point is that classical computers are often necessary to observe how complex patterns emerge from the cellular automatons simple rules. One cannot easily predict how a given classical cellular automaton will behave without stepping through a number of time steps on a computer. Quantum computers could also be used in a similar role for quantum cellular automata as they have been shown to be efficient in the evaluation of path and mutlidimensional integration used in the evolution of quantum cellular automata [88, 89]. In addition so called "Garden of Eden" configurations of classical cellular automata [90], which are arrays of cells that cannot be reached by classical evolution of an update rule, can now be reached quantum mechanically in QCA through quantum tunnelling, although selection rules forbidding certain transitions will still be possible.

Classical cellular automata can be used to perform computations and a select subset of cellular automata can be shown to be capable of universal classical computation. Classical cellular automata can be used to emulate classical Turing machines and these are used in computer science as models of computation. Quantum spin cellular automata and quantum fermionic cellular automata include the standard spin and fermionic model of quantum computation as a subclass. Thus we have a quantum generalization of the relation with quantum cellular automata including quantum computing. As quantum cellular automata don't necessarily have to perform any general computation (other than the one describing the particular QCA action or Hamiltonian), it is easier to construct an experimental realization of a quantum cellular automata using quantum spins or quantum dots than construct a quantum computer. Indeed there are already results on experimental QCA while useful quantum computers are still considered to be far in the future. Nevertheless, universal quantum cellular automata [6], as well their relation with quantum Turing machines (which have been shown to be equivalent to the standard model of quantum computing [91-95]) are of great interest for their relations to theoretical quantum computing.

**VI Conclusions**.

We have generalized the definition of quantum cellular automata to include bosonic, fermionic, supersymmetric and spin quantum systems. We have applied the methods of lattice field theories to the quantization of the bosonoic and fermionic models using the path integral formulation which is well suited to a discrete time formalism. We related a supersymmetric quantum cellular automaton to the string bit models of Thorn and Bergman. This constitutes a new application area for quantum cellular automata to fundamental physics at the smallest length scales. It is at this scale where discrete space time effects could be expected to play a role and these are defining characteristics of cellular automata. Finally we have applied the quantum cellular automata formalism to quantum-dot and quantum computing systems and shown than these are subsets of QCA, in agreement with the notion that quantum cellular automata are theoretically more general than quantum computers and more accessible experimentally.

## Acknowledgements

I wish to acknowledge useful discussions about cellular automata and lattice field theory with Arnold Peskin, Steven Wolfram and Michael Creutz.